\documentclass[epj]{svjour}
\usepackage{latexsym}
\usepackage{epsfig}
\usepackage{mathptmx} 
\DeclareSymbolFont{letters}{OML}{txmi}{m}{it} 
\DeclareMathAlphabet{\mathcal}{OMS}{cmsy}{m}{n} 
\usepackage{amsmath}
\usepackage{amssymb}
\begin{document}
\title{Production of $\eta$ and 3$\pi$ mesons in the $pd \rightarrow ^3$He$X$ reaction at 1360 and 1450 MeV}
\author{K. Sch\"onning \inst{1}, Chr. Bargholtz \inst{2}, M. Bashkanov \inst{3}, M. Berlowski \inst{4}, D. Bogoslawsky \inst{5}, H. Cal\'en \inst{1}, H. Clement \inst{3}, L. Demir\"ors \inst{6}, C. Ekstr\"om \inst{7}, K. Fransson \inst{1}, L. Ger\'en \inst{2}, L. Gustafsson \inst{1}, B. H\"oistad \inst{1}, G. Ivanov \inst{5}, M. Jacewicz \inst{1}, E. Jiganov \inst{5}, T. Johansson \inst{1}, S. Keleta \inst{1}, O. Khakimova \inst{3}, F. Kren \inst{3}, S. Kullander \inst{1}, A. Kupsc \inst{1}, A. Kuzmin \inst{8}, K. Lindberg \inst{2}, P. Marciniewski \inst{1}, B. Morosov \inst{5}, W. Oelert \inst{9}, C. Pauly \inst{6}, H. Petr\'en \inst{1}, Y. Petukhov \inst{5}, A. Povtorejko \inst{5}, W. Scobel \inst{6}, R. Shafigullin \inst{10}, B. Shwartz \inst{8}, T. Skorodko \inst{3}, V. Sopov \inst{11}, J. Stepaniak \inst{4}, P.-E. Tegn\'er \inst{2}, P. Th\"orngren Engblom \inst{2}, V. Tikhomirov \inst{5}, A. Turowiecki \inst{12}, G. J. Wagner \inst{3}, M. Wolke \inst{9}, J. Zabierowski \inst{13}, I. Zartova \inst{2} and J. Z{\l}omanczuk \inst{1}}
%
%
\institute{Dep. of Physics and Astronomy, Uppsala University, Box 516, S-751 20 Uppsala, Sweden 
\and Department of Physics, Stockholm University, S-106 91 Stockholm, Sweden 
\and Physikalisches Institut der Universit\"at T\"ubingen, D-72076 T\"ubingen, Germany
\and So{\l}tan Institute of Nuclear~Studies, PL-00-681 Warsaw, Poland
\and Joint Institute for Nuclear Research, 141980 Dubna, Moscow region, Russia
\and Institut f\"ur Experimentalphysik, Universit\"at Hamburg, D-22761 Hamburg, Germany
\and The Svedberg Laboratory, S-751 21 Uppsala, Sweden
\and Budker Institute of Nuclear Physics, 630 090 Novosibirsk, Russia
\and Institut f\"ur Kernphysik, Forschungszentrum J\"ulich GmbH, D-52425 J\"ulich, Germany
\and Moscow Engineering Physics Institute, Moscow, Russia
\and Institute of Theoretical and Experimental Physics, Moscow, Russia
\and Institute of Experimental Physics, University of Warsaw, PL-00-681 Warsaw, Poland
\and Soltan Institute of Nuclear Studies, PL-90-950 Lodz, Poland
}
\date{Received: date / Revised version: date}
%
\clearpage
\newpage
\abstract{
The cross sections of the $pd \rightarrow ^3$He$\,\eta$, $pd \rightarrow ^3$He$\,\pi^0\pi^0\pi^0$ and $pd \rightarrow ^3$He$\,\pi^+\pi^-\pi^0$ reactions have been measured at the beam kinetic energies $T_p$ = 1360 MeV and $T_p$ = 1450 MeV using the CELSIUS/WASA detector setup. At both energies, the differential cross section $\frac{d\sigma}{d\Omega}$ of the $\eta$ meson in the $pd \rightarrow ^3$He$\,\eta$ reaction shows a strong forward-backward asymmetry in the CMS. The ratio between the $pd \rightarrow ^3$He$\,\pi^+\pi^-\pi^0$ and $pd \rightarrow ^3$He$\,\pi^0\pi^0\pi^0$ cross sections has been analysed in terms of isospin amplitudes. The reconstructed invariant mass distributions of the $\pi\pi$, $^3$He$\pi$ and $^3$He$2\pi$ systems provide hints on the role of nucleon resonances in the $3\pi$ production process.
} 
\authorrunning{K.~Sch\"onning \textsl{et al.}}
\titlerunning{Production of $\eta$ and 3$\pi$ mesons in $pd \rightarrow ^3$He$X$}
\maketitle
\section{Introduction}
\label{intro}
The $pd \rightarrow  \mbox{$^3$He + $X$}$ reaction has  long been used
to  study the  production of  charged and  neutral mesons  and mesonic
systems.  Studying reactions  with  $^3$He in  the  final state  gives
insight  in the  reaction  dynamics involving  three  nucleons and  in
meson-nucleon final state interactions.

The  $pd   \rightarrow  \mbox{$^3$He$\eta$}$  reaction   has  been  of
particular  interest.  Several  studies near  the  kinematic threshold
\cite{berger,mayer,mersmann,smyrski,adam}, where  mostly $s$-waves are
involved in the production process, show a threshold enhancement. This
enhancement  has been interpreted  as an  indication of  a quasi-bound
$^3$He$\eta$ nuclear state  \cite{wilkin1}. Measurements of the $\eta$
angular  distribution at  slightly higher  energies  from PROMICE/WASA
\cite{bilger}  and  ANKE  \cite{rausmann}  indicate  the  presence  of
$p$-waves at an excess energy of $Q\approx$20 MeV, while at $Q\approx$
40 MeV even higher partial waves are required in order to describe the
data. The angular distributions from Refs. \cite{bilger,rausmann} have
a  strong forward-backward  asymmetry with  a backward  suppression, a
maximum at  $\cos\theta_{\eta}^*$ $\approx$ 0.5 and  a forward plateau
or dip.  At slightly overlapping  excess energies, there are data from
GEM  \cite{betigeri} and  Saturne \cite{banaigs}  which  disagree with
the\\ PROMICE/WASA  and ANKE  results. At high  energies ($Q$  $>$ 120
MeV), the data bank is scarce. Backward production of $\eta$ mesons in
$pd \rightarrow \mbox{$^3$He$\eta$}$ was  studied at 17 different beam
energies  at the SPES  IV spectrometer  \cite{Berthet}.  Parts  of the
$\eta$ angular distribution  at $T_p$ = 1450 MeV  was measured by SPES
III  \cite{Kirchner}.   The  CELSIUS/WASA collaboration  has  recently
studied the $pd \rightarrow  \mbox{$^3$He$\eta$}$ reaction at two beam
energies,  \textsl{i.e.}   $T_p$=1450 MeV  and  $T_p$=1360 MeV,  which
correspond to  excess energies of  252 MeV and 299  MeV, respectively.
The differential cross section was measured in the backward hemisphere
and at forward angles. At  $T_p$=1450 MeV, the backward points overlap
with  those from  Ref. \cite{Kirchner}.   The angular  distribution at
$T_p$=1360  MeV obtained with  CELSIUS/WASA is  the first  measured at
this energy.

The direct production of three pions, \textit{i.e.} pions which do not
originate  from \textit{e.g.}  $\omega$ or  $\eta$ decay,  has  so far
received little theoretical and  experimental attention. In the isobar
model  discussed  in Ref.  \cite{3pi1},  three-pion production  should
proceed \textit{via}  an excitation of  one or two  baryon resonances,
like  $\Delta(1232)$  or  the  Roper $N^*(1440)$,  followed  by  their
subsequent decays.  Three-pion production in  proton-proton collisions
was studied  at high energies  \cite{alexander,colleraine,almeida} and
at lower  energies by CELSIUS/WASA  \cite{pauly}. In the  latter work,
the  ratio between  $\sigma(pp  \rightarrow pp\,\pi^+\pi^-\pi^0)$  and
$\sigma(pp   \rightarrow   pp\,\pi^0\pi^0\pi^0)$   was  measured   and
discussed in terms of isospin amplitudes. The ratio was measured to be
$6.3\pm  0.6\pm 1.0$  which suggests  that the  $N^*(1440) \rightarrow
\Delta\pi$ being the  leading part of the reaction  mechanism, in line
with the isobar model presented in Ref. \cite{3pi1}.

In the $pd \rightarrow ^3$He$\pi\pi\pi$ case, it is straightforward to
show  that the total  cross sections  expressed in  isospin amplitudes
$M_{T_{3\pi}}$ are
\begin{equation}
\sigma(pd \rightarrow \mbox{$^3$He$\,\pi^+\pi^-\pi^0$})\propto \frac{2}{15}|M_{1}|^2 + \frac{1}{6}|M_{0}|^2 + \mathrm{cross ~terms}
\label{eq:3ppi1}
\end{equation}

\begin{equation}
\sigma(pd \rightarrow \mbox{$^3$He$\,\pi^0\pi^0\pi^0$})  \propto \frac{1}{30}|M_{1}|^2 
\label{eq:3ppi2}
\end{equation}
where $T_{3\pi}$ denotes the isospin of the three pions. In the simple
statistical approach as outlined by Fermi \cite{fermi}, all amplitudes
in eq.   \ref{eq:3ppi1} and eq.  \ref{eq:3ppi2} are put equal  and the
cross terms  are neglected. Though not  justified, this simplification
enables a  rough comparison between  two channels for which  no other,
more realistic, model exists. The cross section ratio then becomes

\begin{equation}
\frac{\sigma(pd \rightarrow \mbox{$^3$He$\,\pi^+\pi^-\pi^0$})}{\sigma(pd \rightarrow \mbox{$^3$He$\,\pi^0\pi^0\pi^0$})} = 9
\label{eq:ratio}
\end{equation}

If $M_{0}$ is put to  0, the ratio becomes 4. In
this work,  the ratio has  been estimated experimentally at  1360 MeV,
which corresponds  to an excess energy  of $Q$ = 395  MeV for $3\pi^0$
and  $Q$ =  386  MeV for  $\pi^+\pi^-\pi^0$,  and at  1450 MeV,  which
corresponds  to $Q$  = 441  MeV for  $3\pi^0$ and  $Q$ =  432  MeV for
$\pi^+\pi^-\pi^0$.

Multipion production is also interesting since it constitutes the most
important  background to  other  meson production  reactions like  $pd
\rightarrow   ^3$He$\,\eta$,  $pd   \rightarrow   ^3$He$\,\omega$, $pd
\rightarrow ^3$He$\,\Phi$ and \\$pd \rightarrow ^3$He$\,\eta\pi^0$.

This paper is organised as follows: in the next section, the reader is
introduced to  the CELSIUS/WASA experiment.  In section \ref{sec:eta},
the  measurement  of the  $pd  \rightarrow  ^3$He$\,\eta$ reaction  is
presented  and  in  section  \ref{sec:multipi},  the  $pd  \rightarrow
^3$He$\,\pi^0\pi^0\pi^0$ and\\$pd \rightarrow ^3$He$\,\pi^+\pi^-\pi^0$
reactions are studied and compared. Finally the results are summarised
and discussed in section \ref{sec:sumall}.

\section{The CELSIUS/WASA experiment}
\label{sec:celsiuswasa}

The measurements  were carried out  at the The Svedberg  Laboratory in
Uppsala, Sweden. The WASA  detector \cite{Zabierowski} was, until June
2005,  an  integrated  part  of  the  CELSIUS  storage  ring.  In  the
measurements   presented   here,  a   target   of  deuterium   pellets
\cite{Ekstrom,Nordhage}  was  used,  designed  for a  4$\pi$  detector
geometry and high luminosity. 

The  $^3$He   ions  were  detected   in  the  Forward   Detector  (FD)
\cite{calen},   covering    polar   angles   from    3$^{\rm{o}}$   to
18$^{\rm{o}}$.   The  FD consists  of  the  Window  Counter (FWC)  for
triggering, the  Proportional Chamber for  precise angular information
(FPC), the Trigger Hodoscope (FTH) for triggering and offline particle
identification and the Range  Hodoscope (FRH) for energy measurements,
particle  identification  and  triggering.   Mesons  and  their  decay
products  are mainly  detected  in the  central  detector (CD),  which
consists  of the Plastic  Scintillating Barrel  (PSB), the  Mini Drift
Chamber (MDC) and the Scintillating Electromagnetic Calorimeter (SEC).
Charged particles,  mainly pions, are discriminated  from neutral ones
by  their signals  in the  PSB, that  also provides  azimuthal angular
information and  covers a polar angular range  from $24^\mathrm{o}$ to
$159^\mathrm{o}$.  The  momenta of charged particles  are extracted by
tracking in a magnetic field in  the MDC.  The SEC measures angles and
energies of  photons from  meson decays and  covers polar  angles from
$20^\mathrm{o}$  to $169^\mathrm{o}$.  

A special  trigger was developed to  select events with  $^3$He in the
final  state, based  on the  condition  that $^3$He  events give  high
energy deposit  in the FWC and that  hits detected by the  FWC and the
consecutive  detectors  FTH and  FRH  should  match  in the  azimuthal
angle.  It was  carefully checked  in  the offline  analysis that  the
energy deposit  thresholds were set sufficiently low  to accept $^3$He
ions in the full energy range, \textit{i.e.} giving an unbiased $^3$He
sample.   

In the offline  analysis, the $^3$He ions are identified  in the FD by
first  obtaining  a  preliminary  particle identity  (PID)  using  the
$\Delta E$-$E$-method.  In short, we  compare the light output  in the
detector layer  where the  particle stops to  the light output  in the
preceding  layer.   The  $\chi^2$  of  the PID  hypothesis  was  then
calculated by  comparing the measured energy deposits  in all detector
layers   traversed  by   the   particle  to   the  calculated   energy
deposits. Particle hypotheses giving  a $\chi^2$ larger than a certain
maximum     value     were      rejected.     For     details,     see
Ref. \cite{karinthesis,karin3}.

\section{The $pd \rightarrow ^3$He$\,\eta$ reaction}
\label{sec:eta}

The  WASA data collected  at $T_p  = 1360$  MeV and  $T_p =  1450$ MeV
correspond to excess energies  $Q$ = 252 MeV and $Q$ =  299 MeV and to
$\eta$ CM momenta  of $p_{\eta}^*$ = 516 MeV/c  and $p_{\eta}^*$ = 568
MeV/c. Here and in the  following, the star indicates that a kinematic
variable is in the CM system.

The WASA Forward Detector does not cover the entire $^3$He phase space
in  the   \mbox{$pd  \rightarrow  ^3$He$\,\eta$}   reaction  at  these
energies. The maximum  emission angle of the $^3$He  in the laboratory
system is 18.5$^\mathrm{o}$ at $T_p$=1360 MeV and 19.6$^\mathrm{o}$ at
$T_p$=1450 MeV and the FD  only covers angles up to 18.0$^\mathrm{o}$.

\noindent  In  figure   \ref{fig:acc_eta},  the  acceptances  at  both
energies  are  shown  as  a function  of  $\cos\theta^*_{\eta}$,  when
constraints  optimised for  $\eta \rightarrow  \gamma\gamma$ selection
(see section \ref{sec:eta2gamma}) are applied. The acceptance drops at
high and  low angles  due to $^3$He  ions emitted at  small laboratory
angles,  $\theta_{^3He}<$~3$^\mathrm{o}$.  The   middle  hole  in  the
acceptance  is   caused  by  $^3$He  ions  emitted   at  large  angles
$\theta_{^3He}>$~18$^\mathrm{o}$. The acceptance drops at $\cos\theta^*_{\eta} \approx -0.75$ (1360 MeV) and $\cos\theta^*_{\eta} \approx -0.55$ (1450 MeV) are caused by $^3$He ions stopping between two layers of the FRH.

\subsection{Event selection}

The  three main  decay  channels of  the  $\eta$, \textit{i.e.}  $\eta
\rightarrow    \gamma\gamma$    (BR=39.3$\%$),    $\eta    \rightarrow
\pi^0\pi^0\pi^0$ (BR=32.6$\%$)  and $\eta \rightarrow \pi^0\pi^+\pi^-$
(BR=22.7$\%$)  have  all been  separated  and  studied  with the  WASA
setup. In this work, we focus on $\eta \rightarrow \gamma\gamma$ since
it  provides a  clean sample  with good  statistics.  The simultaneous
study  of  $\eta \rightarrow  \pi^0\pi^0\pi^0$  and $\eta  \rightarrow
\pi^+\pi^-\pi^0$ allow valuable cross checks of the results.

\subsubsection{$pd \rightarrow \mbox{$^3$He$\eta, \eta \rightarrow \gamma\gamma$}$}
\label{sec:eta2gamma} 
In this case  all final state particles -- one  $^3$He and two photons
-- can   be  measured   with  good   acceptance.  We   thus   have  an
over-constrained measurement and thereby, we  can check if an event is
consistent with  the expected kinematics. This  reduces the background
significantly and gives a clean sample.

The criteria  for $\eta \rightarrow \gamma\gamma$  selection are given
in table \ref{tab:eta_ggt}. Assuming phase space production, they give
an acceptance of 20\% at 1360 MeV and 14\% at 1450 MeV.

\begin{table}
\begin{center}
\caption{The   constraints   applied   for  selection   of   \mbox{$pd
    \rightarrow       ^3\mathrm{He}\eta,       \,\eta      \rightarrow
    \gamma\gamma$}.  The  angle $\theta_{(\gamma\gamma)mm(^3He)}$  $<$
  $20^{\mathrm{o}}$  refers difference  between the  direction  of the
  $\gamma\gamma$ system and the missing momentum of the $^3$He.}
\begin{tabular}{l}
\hline
 $^3$He giving signal in the FPC and stopping in the FRH \\
 $\geqslant$ 2 photons in the SEC with $E_{\gamma}$ $>$ 20 MeV\\
 one $\gamma\gamma$-combination fulfilling\mbox{$|IM(\gamma\gamma)-m_{\eta}|< 150~\rm{MeV}/c^2$} \\
 $MM^2(^3$He$\gamma\gamma)$ $<$ 10000 $(\rm{MeV}/c^2)^2$ \\
 $\theta_{(\gamma\gamma)mm(^3He)}$ $<$ $20^{\mathrm{o}}$ \\
 no overlapping hits in the PSB and the SEC \\
 160$^\mathrm{o}<|\phi_{lab}(^3$He$)-\phi_{lab}(\gamma\gamma)|<200^\mathrm{o}$ \\
\hline
\end{tabular}
\label{tab:eta_ggt}
\end{center}
\end{table}

\begin{figure}
\resizebox{0.5\textwidth}{!}{%
\includegraphics{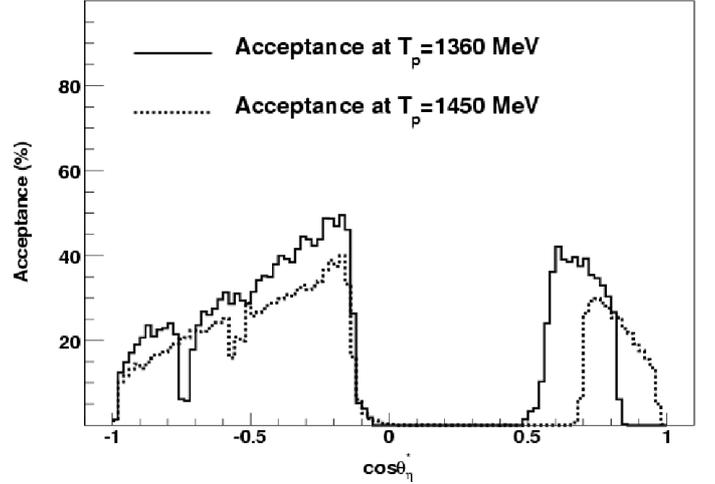}
}
\caption{The acceptance of the $pd \rightarrow ^3$He$\,\eta,\,\eta \rightarrow \gamma\gamma$ reaction as a function of $\cos\theta_{\eta}^*$ at $T_p$=1450 MeV (dashed line) and at $T_p$=1360 MeV (solid line).}
\label{fig:acc_eta}       
\end{figure}

\noindent Figure \ref{fig:acc_eta} shows  how the acceptance varies as
a function of $\cos\theta_{\eta}^*$.  The acceptance is limited by the
geometrical coverage of the FD,  by photons missing the CsI modules in
the calorimeter  and by  the efficiency reduction  due to  $^3$He ions
undergoing nuclear interaction before depositing all their energy.

\begin{figure}
\begin{center}
\resizebox{0.5\textwidth}{!}{%
\includegraphics{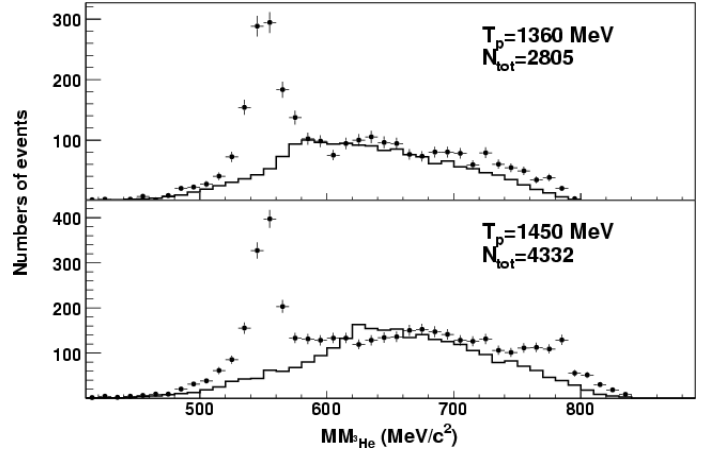}
}
\caption{The  upper panel shows  the WASA  data sample  fulfilling the
  constraints   optimised   for   selection   of   $\eta   \rightarrow
  \gamma\gamma$ at  1360 MeV  and the lower  panel shows the  1450 MeV
  case.   The solid  line histograms  show Monte  Carlo  simulated $pd
  \rightarrow  \mbox{$^3$He$\,\pi^0\pi^0$}$ data fulfilling  the given
  constraints. The  spectra are not  corrected for acceptance  and the
  background simulations are scaled to fit the data.}
\label{fig:eta_2g}       
\end{center}
\end{figure}

The upper  panel of figure  \ref{fig:eta_2g} shows the  $^3$He missing
mass  for all events  fulfilling the  constraints optimised  for $\eta
\rightarrow  \gamma\gamma$  selection at  $T_p$=1360  MeV. The  bottom
panel  shows  the  $T_p$=1450  MeV  case.   Phase  space  Monte  Carlo
simulations  of  the main  background  channel, \mbox{$pd  \rightarrow
  \mbox{$^3$He$\,\pi^0\pi^0$}$}, are also shown, normalised to fit the
data. They  reproduce the background  in the experimental  data fairly
well,  except  for an  enhancement  at  high  $^3$He missing  mass  at
$T_p$=1450   MeV   which   is   caused  by\\   \mbox{$pd   \rightarrow
  \mbox{$^3$He$\,\omega$},  \,\omega \rightarrow  \pi^0\gamma$} events
that accidentally satisfy the  criteria. Assuming phase space $2\pi^0$
production give an acceptance of 3.6$\%$ at $T_p$=1360 MeV and 4.0$\%$
at   $T_p$=1450   MeV.   Other  reactions,   \textit{e.g.}   \mbox{$pd
  \rightarrow \mbox{$^3$He$\,\pi^0\pi^0\pi^0$}$}, were found to give a
negligible  contribution   to  the  $\eta   \rightarrow  \gamma\gamma$
background.

\subsubsection{$pd \rightarrow \mbox{$^3$He$\eta, \eta \rightarrow \pi^0\pi^0\pi^0$}$}
\label{sec:eta_3pi0}
In this  case we  need six  photons from the  three $\pi^0$  decays in
order  to  indentify  the  events.  The  Scintillator  Electromagnetic
Calorimeter (SEC)  has a small ``hole''  in the backward  part and one
large   in    the   forward    part,   where   the    photons   escape
undetected.  Therefore,  in  most $\eta  \rightarrow  \pi^0\pi^0\pi^0$
events at least one, but  often several, photons escape detection. The
acceptance is  therefore significantly  reduced compared to  the $\eta
\rightarrow \gamma\gamma$ case.

The constraints optimised for $\eta \rightarrow \pi^0\pi^0\pi^0$ selection are given in table \ref{tab:eta_3pi0t}.

\begin{table}
\begin{center}
\caption{The constraints applied for selection of \mbox{$pd \rightarrow ^3\mathrm{He}\eta, \,\eta \rightarrow \pi^0\pi^0\pi^0$.}}
\begin{tabular}{l}
\hline
 $^3$He giving signal in the FPC and stopping in the FRH \\
 $\geqslant$ 6 photons in the SEC with $E_{\gamma}$ $>$ 20 MeV\\
 one $\gamma\gamma$-combination fulfilling \mbox{$|IM(\gamma\gamma)-m_{\pi^0}|< 50~\rm{MeV}/c^2$} \\
 two other $\gamma\gamma$-combinations fulfilling\mbox{$|IM(\gamma\gamma)-m_{\pi^0}|< 60~\rm{MeV}/c^2$} \\
 $MM^2(^3$He$6\gamma)$ $<$ 20000 $(\rm{MeV}/c^2)^2$ \\
 no overlapping hits in the PSB and the SEC \\
 \hline
\end{tabular}
\label{tab:eta_3pi0t}
\end{center}
\end{table}

\noindent  Assuming  phase  space   production,  this  gives  a  total
acceptance  of  5.7$\%$  at  $T_p$=1360  MeV  and  3.6$\%$  $T_p$=1450
MeV.  The  main background  channel  is  direct \mbox{$pd  \rightarrow
  ^3$He$\,\pi^0\pi^0\pi^0$} production. At  high missing masses, there
is     also    a     contribution    from\\ \mbox{$pd    \rightarrow
  ^3$He$\,\pi^0\pi^0\pi^0\pi^0$}  production, which will  be discussed
in  section   \ref{sec:3pi0}.  The  acceptance   for  direct  3$\pi^0$
production at  $T_p$=1360 MeV is  11.7$\%$ and 10.3$\%$  at $T_p$=1450
MeV, if phase space production is assumed.

\begin{figure}
\begin{center}
\resizebox{0.5\textwidth}{!}{%
\includegraphics{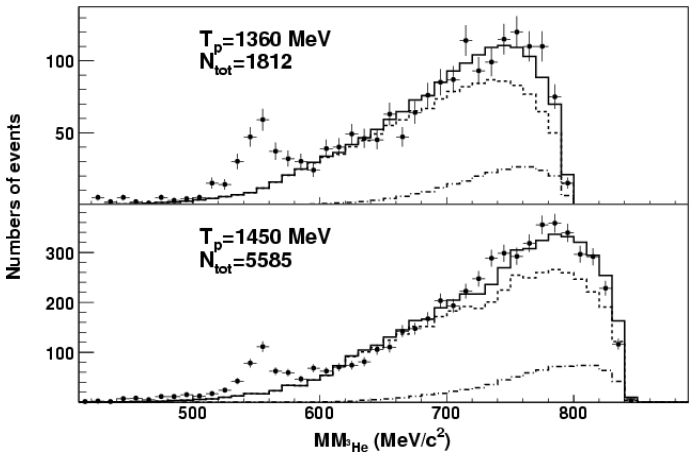}
}
\caption{The  upper panel  shows the  data fulfilling  the constraints
  optimised  for selection  of $\eta  \rightarrow  \pi^0\pi^0\pi^0$ at
  1360 MeV  and the lower panel  shows the 1450 MeV  case.  The dotted
  line   histograms  show  Monte   Carlo  simulated   $pd  \rightarrow
  \mbox{$^3$He$\,\pi^0\pi^0\pi^0$}$   data    fulfilling   the   given
  constraints, the  dashed-dotted histogram simulated  $pd \rightarrow
  \mbox{$^3$He$\,\pi^0\pi^0\pi^0\pi^0$}$ data  and the solid  line the
  sum  of  3$\pi^0$  and  4$\pi^0$  production. The  spectra  are  not
  corrected for  acceptance and the background  simulations are scaled
  to fit the data.}
\label{fig:eta_3pi0}       
\end{center}
\end{figure}

The upper panel of  figure \ref{fig:eta_3pi0} shows the $^3$He missing
mass  for all events  fulfilling the  constraints optimised  for $\eta
\rightarrow \pi^0\pi^0\pi^0$  at $T_p$=1360  MeV and the  bottom panel
shows the same but for $T_p$=1450 MeV.

\subsubsection{$pd \rightarrow \mbox{$^3$He$\eta, \eta \rightarrow \pi^+\pi^-\pi^0$}$}
\label{sec:eta_pipimpi0}

The  criteria  optimised for  $\eta  \rightarrow \pi^+\pi^-\pi^0$  are
given in  table \ref{tab:eta_pipimpi0t}.  The last one,  requiring the
total energy  deposit in the SEC  to be smaller than  900 MeV, rejects
time-overlapping   events,  \textit{i.e.}  chance   coincidences.  The
selection  criteria  give  altogether   an  acceptance  of  18$\%$  at
$T_{p}=1360$ MeV and 12$\%$ at $T_{p}=1450$ MeV.
\begin{table}
\begin{center}
\caption{The constraints applied for selection of \mbox{$pd \rightarrow ^3\mathrm{He}\eta, \,\eta \rightarrow \pi^+\pi^-\pi^0$.}}
\begin{tabular}{l}
\hline
 $^3$He giving signal in the FPC and stopping in the FRH \\
 $\geqslant$ 2 photons in the SEC with $E_{\gamma}$ $>$ 20 MeV\\
 one $\gamma\gamma$-combination fulfilling\mbox{$|IM(\gamma\gamma)-m_{\pi^0}|< 45~\rm{MeV}/c^2$} \\
 $MM(^3$He$\pi^0)$ $>$ 250 $\rm{MeV}/c^2$ \\
 $\geqslant$ 2 hits in the PSB \\
 $E_{\rm{tot}}(\rm{SEC})$ $<$ 900 MeV \\
 \hline
\end{tabular}
\label{tab:eta_pipimpi0t}
\end{center}
\end{table}

\begin{figure}
\begin{center}
\resizebox{0.5\textwidth}{!}{%
\includegraphics{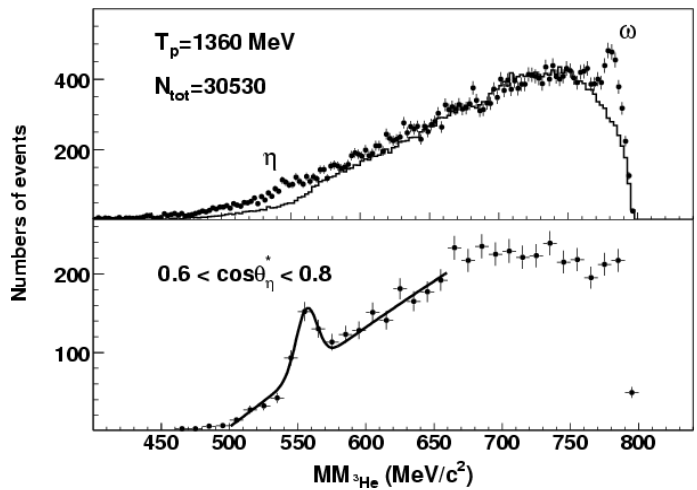}
}
\caption{The upper panel shows all data at $T_p$=1360 MeV that satisfy
  the criteria optimised for $pd \rightarrow \mbox{$^3$He$\,\eta, \eta
    \rightarrow   \pi^+\pi^-\pi^0$}$   selection.   The   solid   line
  represents  Monte Carlo  simulations  of direct  $\pi^{+}\pi^-\pi^0$
  production. These  spectra are not corrected for  acceptance and the
  background simulations are  scaled to fit the data.  The lower panel
  shows    the    same   thing    but    in    the   angular    region
  0.6$<\cos\theta^*_{\eta}<$0.8. The line is the  result of a fit of a
  gaussian peak on top of a polynomial background.}
\label{fig:eta_pipimpi0f}       
\end{center}
\end{figure}

The   main  background   comes   from  nonresonant   $\pi^+\pi^-\pi^0$
production.    The     acceptance    for    the     $pd    \rightarrow
\mbox{$^3$He$\,\pi^+\pi^-\pi^0$}$ reaction  when the given constraints
are  applied and  phase  space  production is  assumed,  is 35$\%$  at
$T_{p}=1360$ MeV and 31$\%$ at $T_{p}=1450$ MeV.

The upper  panel of figure  \ref{fig:eta_pipimpi0f} shows all  data at
$T_p$=1360  MeV that fulfill  the cuts  optimised for  $pd \rightarrow
\mbox{$^3$He$\,\eta, \eta \rightarrow \pi^0\pi^+\pi^-$}$ selection. It
is difficult to separate the $\eta$ events from the background, partly
due  to the  small signal-to-background  ratio and  partly due  to the
broad    $\eta$   peak.    However,    in   individual    regions   in
$\cos\theta^*_{\eta}$, the $\eta$  events appear in a peak  and can be
separated from the background  with reasonable accuracy. An example is
shown in the lower panel of figure \ref{fig:eta_pipimpi0f}. The $\eta$
peak  for the  full $\cos\theta_{\eta}^*$  range, shown  in  the upper
panel of  figure \ref{fig:eta_pipimpi0f},  is broader than  the $\eta$
peak  in an  individual $\cos\theta_{\eta}^*$  interval, shown  in the
lower panel  of figure \ref{fig:eta_pipimpi0f}.  The  broadness of the
peak  in  the full  $\cos\theta_{\eta}^*$  range  is  due to  a  small
dependence of the $\eta$  peak position on $\cos\theta_{\eta}^*$. This
in turn is an effect  of the calibration constants, which are slightly
dependent on energy. This was also observed in Ref. \cite{bilger}, but
there the effect  was much stronger.  Here it  is negligible for small
lab angles $\theta_{^3He}^{lab}$ where  the variation in $T_{^3He}$ is
small. For large $\theta_{^3He}^{lab}$, it gives a contribution to the
systematic uncertainty of $<$ 3 \%.

\subsection{The $\eta$ angular distribution}

The  angular  distributions  were   obtained  by  dividing  the  $\eta
\rightarrow    \gamma\gamma$   data    sample   into    intervals   of
$\cos\theta_{\eta}^*$  where the  acceptance is  smooth  and non-zero.
The  $\eta$  mesons are  identified  by  the  missing mass  method  in
individual  bins  of  $\cos\theta^*_{\eta}$.   The $\eta$  mesons  are
easier to  identify in the  intervals than in the  cumulative spectrum
(compare    the    upper   and    the    lower    panel   of    figure
\ref{fig:eta_pipimpi0f}).    The  number   of  $\eta$   candidates  is
extracted by fitting  Gaussian peak on top of  a polynomial background
(it has been checked that in individual  $\cos\theta^*_{\eta}$ region the
background  has no  discontinuities). This
number  was then  corrected for  acceptance. The  systematic  uncertainty was
estimated by fitting simulated Monte Carlo data of the main background
channel  (in this case $p  d  \rightarrow  ^3$He$\pi^0\pi^0$) and compare the number of $\eta$ events
obtained in this way to the number of $\eta$s obtained from fitting the background to a 
polynomial. The  same  procedure was
repeated  for the $\eta  \to 3\pi$  channels.  It  turns out  that the
agreement in individual  $\cos\theta^*_{\eta}$ regions is good between
the  $\eta  \rightarrow  2\gamma$  and  the  $\eta  \rightarrow  3\pi$
channels.  This  gives confidence that  the cut efficiencies  are well
understood and that our systematic uncertainties are under control.

The  normalisation was achieved  by comparing  data on  backward going
$\eta$  mesons   from  $pd   \rightarrow  ^3$He$\eta$  from   SPES  IV
\cite{Berthet}  and SPES  III \cite{Kirchner}  with  the corresponding
data    from   this    work    using   the    method   described    in
Refs.   \cite{karinthesis}   and   \cite{karin2}.  The   normalisation
uncertainty of the measured cross sections and is 29\% at 1360 MeV and
12\% at 1450  MeV.

\begin{figure}
\begin{center}
\resizebox{0.5\textwidth}{!}{%
\includegraphics{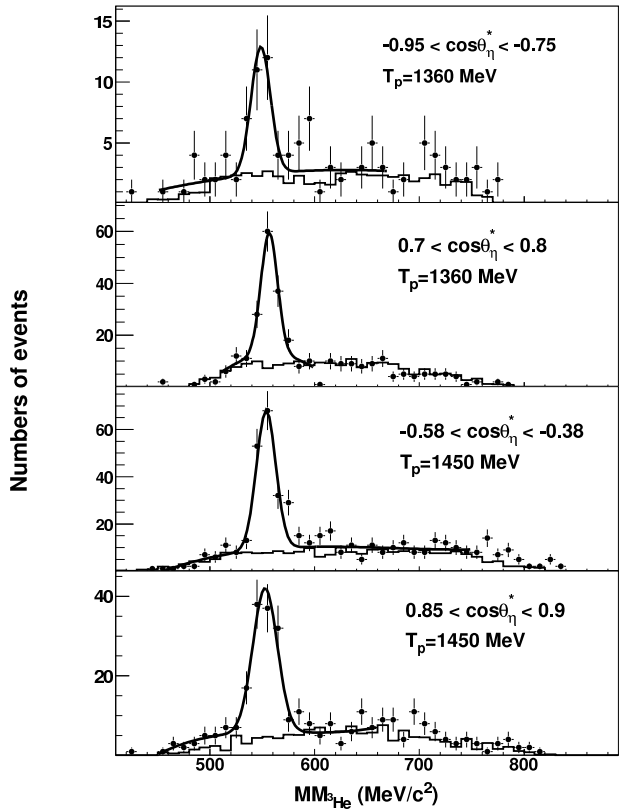}
}
\caption{The  $^3$He missing mass  distribution for  events satisfying
  the  constraints  optimised   for  $\eta  \rightarrow  \gamma\gamma$
  selection     (see      text)     in     given      intervals     of
  $\cos\theta^*_{\eta}$.   These  $MM$-spectra   are   not  acceptance
  corrected  and the  background  simulations are  scaled  to fit  the
  data.}
\label{fig:mm_ang}
\end{center}
\end{figure} 

\begin{figure}

\resizebox{0.5\textwidth}{!}{%
\includegraphics{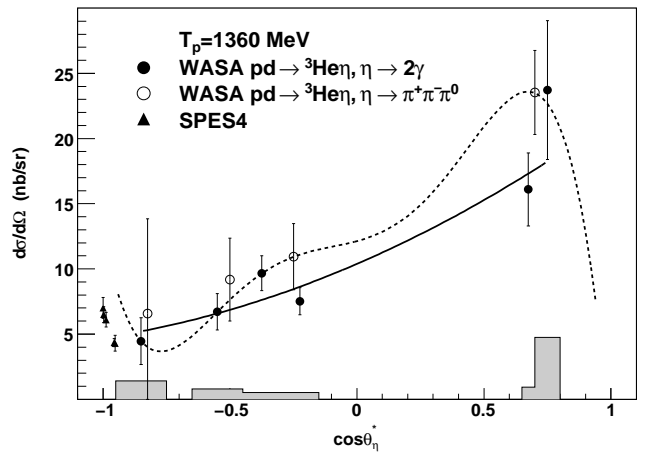}
}
\caption{Angular distribution of the $\eta$  meson in the CM system at
  $T_p$  =  1360 MeV.  The  black  dots are  WASA  data  from the  $pd
  \rightarrow  \mbox{$^3$He$\eta,   \eta  \rightarrow  \gamma\gamma$}$
  channel. The error bars  represent the statistical uncertainties and
  the grey histogram the systematical. The open dots are obtained with
  WASA data  from $pd \rightarrow  \mbox{$^3$He$\eta, \eta \rightarrow
    \pi^0\pi^+\pi^-$}$.   The  black   triangles  are   calculated  by
  interpolating  SPES IV  \cite{Berthet}  data at  $T_p$=1250 MeV  and
  $T_p$=1350 MeV  and SPES III data  at $T_p$=1450 MeV.  The Legendre fit represented by the solid line has been used to calculate the cross section quoted here and fit shown as a dashed line has been used to estimate the systematical uncertainty.}
\label{fig:eta_ang_1360}
\end{figure}

\begin{figure}
\resizebox{0.5\textwidth}{!}{%
\includegraphics{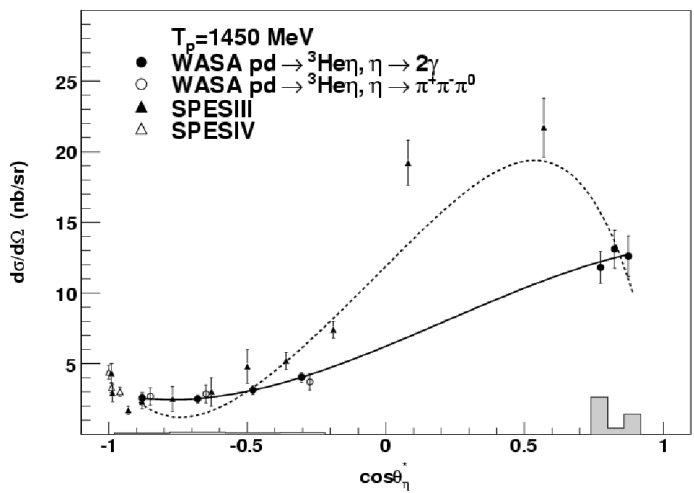}
}
\caption{Angular distribution of the $\eta$  meson in the CM system at
  $T_p$ = 1450 MeV. The black  dots are WASA data from $pd \rightarrow
  \mbox{$^3$He$\eta,  \eta \rightarrow  \gamma\gamma$}$  at $T_p$=1450
  MeV. The error bars  represent the statistical uncertainties and the
  grey  histogram the systematical.  The open  dots are  obtained with
  WASA data  from $pd \rightarrow  \mbox{$^3$He$\eta, \eta \rightarrow
    \pi^0\pi^+\pi^-$}$.  The black  triangles are  data from  SPES III
  \cite{Kirchner}  while   the  open  triangles  come   from  SPES  IV
  \cite{Berthet}. The curves are results of fits of Legendre series to
  the WASA data (solid) and WASA plus SPES III (dashed).}
\label{fig:eta_ang_1450}
\end{figure}

The   resulting   angular    distributions   are   shown   in   figure
\ref{fig:eta_ang_1360}   and    figure   \ref{fig:eta_ang_1450}.   The
systematic  uncertainties are  shown  as a  shaded  histogram in  each
figure.  They  mainly  arise  from  the ambiguity  in  the  background
subtraction, but  there is also  a small contribution from  the energy
dependence    of    the    calibration    constants    (see    section
\ref{sec:eta_pipimpi0}).   The  distributions  at  both  energies  are
highly anisotropic with a sharp forward-backward asymmetry. This is in
line        with       earlier        experiments,       \textit{e.g.}
Refs. \cite{bilger,rausmann,betigeri,banaigs,Kirchner}, where evidence
were found  for several higher  partial waves away from  the threshold
region.  From comparing  SPES III  data with  data from  this  work at
$T_p$=1450 MeV,  which is  done in figure  \ref{fig:eta_ang_1450}, the
conclusion is that either the two data sets are inconsistent, or there
is  a forward  dip that  is  much stronger  than the  dip observed  in
Refs. \cite{bilger,rausmann}.

The  angular  distributions  were  fitted  by  a  series  of  Legendre
polynomials
\begin{equation}
\frac{d\sigma}{d\Omega}(\cos\theta_{\eta}^*)=\displaystyle\sum_{k=0}^{k_{max}}a_{k}P_{k}(\cos\theta_{\eta}^*)
\end{equation}
to  the $\eta  \rightarrow \gamma\gamma$  data points  from  WASA. The
zeroth coefficient  of the Legendre polynomial  gives, when multiplied
with  4$\pi$,  the  total  cross  section. At  1360  MeV  one  obtains
\\ \mbox{$\sigma_{\rm{tot}}$  = $151.6 \pm  9.3 \pm  35.3$ nb}.   In addition,
there is an uncertainty from  the normalisation of 29$\%$. At 1450 MeV
the  total cross  section is  estimated to  be  \mbox{$\sigma_{\rm{tot}}$ =
  $80.9 \pm 3.6  \pm 43.0$ nb}. The normalisation  uncertainty is 12\%
at 1450 MeV.

\section{Multipion production}
\label{sec:multipi}
In   this    section,   we    first   study   the    $pd   \rightarrow
^3$He$\,\pi^0\pi^0\pi^0$   reaction,    then   the   $pd   \rightarrow
^3$He$\,\pi^+\pi^-\pi^0$  reaction  and  finally, the  two  three-pion
reactions are compared.

\subsection{The $pd \rightarrow \mbox{$^3$He$\,\pi^0\pi^0\pi^0$}$ reaction}
\label{sec:3pi0}

The  same selection  criteria  are  used as  for  the $pd  \rightarrow
^3$He$\,\eta,\,\eta \rightarrow \pi^0\pi^0\pi^0$  case, given in table
\ref{tab:eta_3pi0t}  of  section   \ref{sec:eta_3pi0}.   For  the\\ $pd
\rightarrow  ^3$He$\,\pi^0\pi^0\pi^0$ reaction, they  give acceptances
of 11.7$\%$ at 1360 MeV and 10.3$\%$ at 1450 MeV. There may be a large
uncertainty  in the  acceptance of  a reaction  where six  photons are
measured. To estimate this  uncertainty, we assume that the difference
in  the  extracted number  of  $\eta$  mesons  from $\eta  \rightarrow
\gamma\gamma$  and  $\eta  \rightarrow  \pi^0\pi^0\pi^0$  is  entirely
caused by the  ambiguities in the acceptance and  that the uncertainty
is the  same at  both energies. The  uncertainty in the  acceptance is
then estimated to a maximum value of 20\%.

The $^3$He missing mass distributions  at both energies for all events
fulfilling the  constraints are shown in  figure \ref{fig:eta_3pi0} in
section \ref{sec:eta_3pi0}. The  dotted line shows simulated \mbox{$pd
  \rightarrow  \mbox{$^3$He$\,\pi^0\pi^0\pi^0$}$} data  assuming phase
space production. Simulated 3$\pi^0$  data match the experimental data
for low and medium missing masses (except at the $\eta$ peak, which is
expected), but at high MM($^3$He), the matching between data and phase
space Monte Carlo is poor.

It is reasonable to assume a contribution from\\ \mbox{$pd \rightarrow
  \mbox{$^3$He$\,\pi^0\pi^0\pi^0\pi^0$}$},    either    from    direct
production  or   from  production  \textit{via}   $\eta\pi^0$  in  $pd
\rightarrow       \mbox{$^3$He$\,\eta\pi^0,      \eta      \rightarrow
  \pi^0\pi^0\pi^0$}$.  In both reactions,  eight photons  are produced
and   the   acceptance   for   the   selection   criteria   in   table
\ref{tab:eta_3pi0t} is 28\%  at 1360 MeV and 24\% at  1450 MeV. At the
highest energy, the maximum $^3$He emission angle in the lab system is
15$^{\rm{o}}$  in  the  $\eta\pi^0$  case  and  18$^{\rm{o}}$  in  the
$4\pi^0$  case, which  means  that  in both  cases,  the WASA  Forward
Detector covers almost the full  $^3$He phase space. The acceptance is
then nearly independent of the production mechanism.

The $4\pi^0$  distributions obtained from Monte  Carlo simulations are
shown  in the  dashed-dotted line  histograms in  the upper  and lower
panel  of  figure \ref{fig:eta_3pi0}.  Adding  the contributions  from
3$\pi^0$  and 4$\pi^0$  together gives  the solid  line  histograms in
figure  \ref{fig:eta_3pi0}. We obtain  $N_{4\pi^0}$ =  250 at  $T_p$ =
1360 MeV and $N_{4\pi^0}$ = 800 at $T_p$ = 1450 MeV.

The cross  section of the  $pd \rightarrow \mbox{$^3$He$\,\eta\pi^0$}$
reaction at $T_p$ = 1450 MeV has been measured to $\sigma_{\rm{tot}} =
23.6 \pm 1.6 \pm 2.2$ nb  $\pm 14\%$ by studying the $\eta \rightarrow
\gamma\gamma$ decay \cite{karinetapi}.

From    the     known    cross    section     of    $pd    \rightarrow
\mbox{$^3$He$\,\eta\pi^0$}$  reaction   at  $T_p$  =   1450  MeV  (see
Ref.  \cite{karinetapi}) and  from  the acceptance  and the  branching
ratio of  $\eta \rightarrow  \pi^0\pi^0\pi^0$, the expected  number of
$pd    \rightarrow    \mbox{$^3$He$\,\eta\pi^0,    \eta    \rightarrow
  \pi^0\pi^0\pi^0$}$  events  is  calculated  to 700  $\pm$  80.  This
explains almost fully the $N_{4\pi^0}$ =  800 and it is clear that the
cross section of direct 4$\pi^0$ production must be very small at 1450
MeV.

Subtracting the fitted $4\pi^0$ and $\eta\pi^0$ distributions from the
experimental data gives $N_{3\pi^0}$ = 1400 and $N_{3\pi^0}$ = 4500 at
$T_p$ = 1360 MeV and  $T_p$ = 1450 MeV, respectively. This corresponds
to  $3\pi^0$ cross sections  of 180  nb and  115 nb.   The statistical
uncertainty of $N_{3\pi^0}$  is given by the square  root of the total
number of events  before the subtraction and is equal  to 3\% (2\%) at
$T_p$=1360 MeV ($T_p$=1450 MeV).

It is also possible that part of the deviation from the 3$\pi^0$ phase
space  curve  in figure  \ref{fig:eta_3pi0}  is  due  to a  production
mechanism  that differs  from  phase space  production.  This will  be
discussed later  in this paper.  However, at least at  $T_p$=1450 MeV,
the      expected      contribution      from     $pd      \rightarrow
\mbox{$^3$He$\,\eta\pi^0,\,\eta \rightarrow 3\pi^0$}$ explain the data
well and the remaining excess  of events at high $^3$He missing masses
gives a small contribution to the systematic uncertainty.

At  $T_p$=1360 MeV, it  is difficult  to say  with certainty  that the
excess of  events at  high $MM(^3\mathrm{He})$ in  the upper  panel of
figure   \ref{fig:eta_3pi0}  are   not   directly  produced   $3\pi^0$
events.    The    cross     section    of    the    $pd    \rightarrow
\mbox{$^3$He$\,\eta\pi^0$}$ reaction is not known, and it is therefore
unclear whether a significant contribution from this reaction is to be
expected.  However,  the  mixture  of  $3\pi^0$  and  $4\pi^0$  events
reproduces  the experimental distributions  also at  $T_p$ =  1360 MeV
very well and it is therefore reasonable to assume a contribution from
$4\pi^0$  production,  either  from  direct  production  or  from  the
subsequent $\eta$  decay in $\eta\pi^0$ production.  We therefore take
the 3$\pi^0$  cross section of  180 nb, calculated when  assuming that
the deviation from the 3$\pi^0$  curve at large $MM(^3$He$)$ in figure
\ref{fig:eta_3pi0} come from 4$\pi^0$ production, as the most reliable
one. The excess of events is  treated as a systematic uncertainty. By
assuming   that   all   events   in   the  upper   panel   of   figure
\ref{fig:eta_3pi0} that do not come from $pd \rightarrow ^3$He$\,\eta,
\eta  \rightarrow  3\pi^0$  are  directly  produced  3$\pi^0$  events,
$N_{3\pi^0}$ becomes 1650 which corresponds  to a cross section of 212
nb. The systematic uncertainty is then taken as the difference between
the cross sections calculated  in two different ways, \textit{i.e.} 32
nb.  We  assume   that  the  uncertainty  is  symmetric.   This  is  a
conservative method of estimating the systemtaic uncertainty and other
systematic contributions, \textit{e.g} variation in the acceptance due
to reaction mechanism, should be  well within the error bars estimated
in this way.

We  can  also  give  a  rough  upper  limit  of  the  $pd  \rightarrow
^3$He$\,\eta\pi^0$  at 1360  MeV, which  will  be useful  in the  next
section. Assuming  that all  the $N_{4\pi^0}$ =  250 events  come from
$\eta\pi^0$ production, the $\sigma(pd \rightarrow ^3$He$\,\eta\pi^0)$
would be 42 nb.

The  total   cross  section   of  $3\pi^0$  production   then  becomes
$\sigma_{3\pi^0} = 180 \pm 6 \pm 49$ nb $\pm 29\%$ at $T_p$ = 1360 MeV
and $\sigma_{3\pi^0} = 115 \pm 3 \pm 23$ nb $\pm 12\%$ at $T_p$ = 1450
MeV. The  first error is  statistical, the second is  systematical and
includes  uncertainties  from  background  and  acceptance.  The  last
uncertainty comes from the normalisation.

Background     from    quasi-free     reactions     $pp    \rightarrow
pp\,\pi^0\pi^0\pi^0$ with a  proton misidentified as a $^3\mathrm{He}$
is expected to be negligible. The probability that an event from
a  reaction  with   $p$  or  $d$  in  the   final  state  instead  of
$^3\mathrm{He}$ would survive the constraints, is smaller than 0.001\%.

Invariant mass distributions  of the final state particles  in the $pd
\rightarrow    ^3$He$\,\pi^0\pi^0\pi^0$   reaction    give   important
information about the production mechanism. Deviation from phase space
can give  hints about  \textit{e.g.} intermediate resonances.  In this
work we have studied the $2\pi^0$-system, the $^3$He$\pi^0$-system and
the  $^3$He$2\pi^0$-\\system. When  studying the  invariant mass  of two
pions it is more convenient to instead reconstruct the missing mass of
the $^3$He and the third pion, here denoted $MM(^3$He$\pi^0)$. This is
because the $^3$He  is measured in the FD  with higher resolution than
the pions, which are measured in the CD.

\begin{figure}
\begin{center}
\includegraphics[width=\columnwidth,clip]{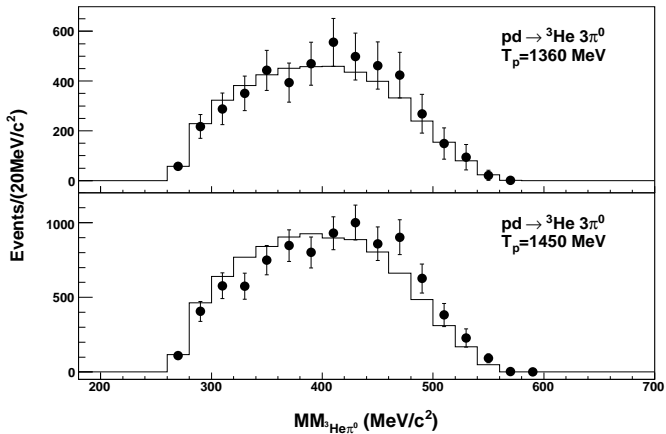}
\caption{The  missing  mass  of  the  $^3$He$\pi^0$-system,  which  is
  equivalent to the invariant  mass of the $2\pi^0$-system. The points
  represent  background  subtracted   and  acceptance  corrected  data
  satisfying  the criteria  given in  the  text and  $600$ MeV$/c^2  <
  MM(^3$He$) <  700$ MeV$/c^2$. The solid histogram  shows phase space
  Monte  Carlo simulations  of  $3\pi^0$ production.  The upper  panel
  shows the $T_p$ =  1360 MeV case and the lower the  $T_p$ = 1450 MeV
  case.}
\label{fig:im2p_3pi0}       
\end{center}
\end{figure}

In order  to avoid an event sample  with a lot of  background from the
$pd \rightarrow ^3$He$\,\eta$  and $pd \rightarrow ^3$He$\,\eta\pi^0$,
events which fulfill the condition \mbox{$600$ MeV$/c^2 < MM(^3$He$) <
  700$   MeV$/c^2$}   are    selected.   $MM(^3$He$\pi^0)$   is   then
reconstructed for these events.  In this event sample, there will be a
small  contribution (a  few  percent) from  the \mbox{$pd  \rightarrow
  ^3$He$\,\eta\pi^0, \eta  \rightarrow 3\pi^0$} reaction  and the data
will  therefore be subtracted  by the  expected amount  of $\eta\pi^0,
\eta \rightarrow  3\pi^0$ events, obtained from  simulations. The data
are then corrected for acceptance.  The results are shown in the upper
and lower  panel of  figure \ref{fig:im2p_3pi0}. The  points represent
the background subtracted and  acceptance corrected data and the solid
histogram phase  space simulated $3\pi^0$ data.  The experimental data
follow phase space well.

\begin{figure}
\begin{center}
\includegraphics[width=\columnwidth,clip]{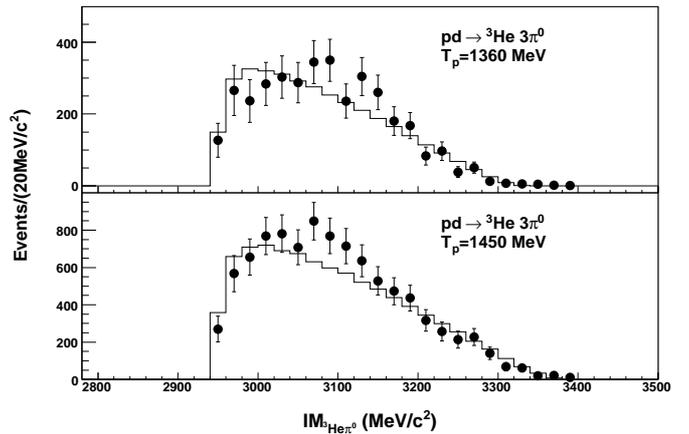}
\caption{The  invariant mass of  the $^3$He$\pi^0$-system.  The points
  represent  background  subtracted  and  acceptance  corrected  data,
  fulfilling the  constraints given in  the text and $600$  MeV$/c^2 <
  MM(^3$He$) <  700$ MeV$/c^2$. The solid histogram  shows phase space
  Monte  Carlo simulations  of  $3\pi^0$ production.  The upper  panel
  shows the $T_p$ =  1360 MeV case and the lower the  $T_p$ = 1450 MeV
  case.}
\label{fig:imhp_3pi0}       
\end{center}
\end{figure}

The invariant mass of the $^3$He$\pi^0$-system, $IM(^3$He$\pi^0)$, has
also  been reconstructed.  The small  background from  $\eta\pi^0$ was
subtracted in  the same way as  in the $MM(^3$He$\pi^0)$  case and the
data was then corrected for  acceptance. The result is shown in figure
\ref{fig:imhp_3pi0}. Here the data disagree with phase space. There is
an enhancement  with respect to  phase space centered  around $\approx
3090$   MeV$/c^2$,   which  roughly   equals   the   sum   $2  m_p   +
M_{\Delta(1232)}$.  This  may  indicate  a  single
$\Delta(1232)$ excitation in the production process.

\begin{figure}
\begin{center}
\includegraphics[width=\columnwidth,clip]{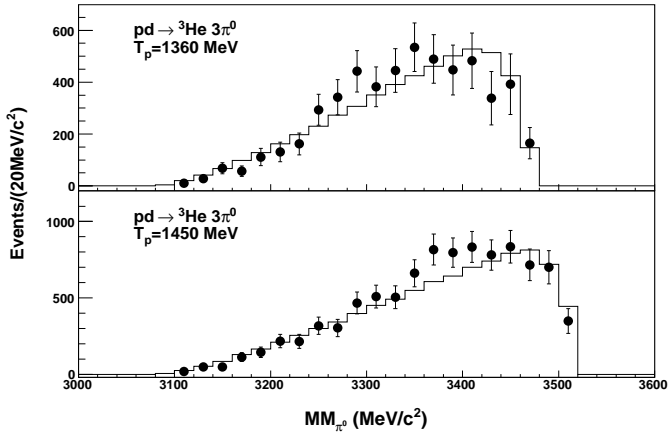}
\caption{The missing mass  of the $\pi^0$, which is  equivalent to the
  invariant  mass of the  $^3$He$2\pi^0$-system. The  points represent
  background subtracted  and acceptance corrected  data fulfilling the
  constraints given in the text and $600$ MeV$/c^2 < MM(^3$He$) < 700$
  MeV$/c^2$.  The  solid  histogram  shows  phase  space  Monte  Carlo
  simulations of $3\pi^0$ production.  The upper panel shows the $T_p$
  = 1360 MeV case and the lower the $T_p$ = 1450 MeV case.}
\label{fig:mmpi0_3pi0}       
\end{center}
\end{figure}

Finally,  the  invariant  mass  of  the  $^3$He$\pi^0\pi^0$-system  is
studied. This is done by reconstructing the missing mass of one of the
$\pi^0$  mesons.  In  this   way,  the  resolution  is  improved.  The
experimental   data  were   background  subtracted,   using  simulated
\mbox{$pd  \rightarrow  ^3$He$\,\eta\pi^0,  \eta \rightarrow  3\pi^0$}
data,  and  acceptance  corrected.  The  result  is  shown  in  figure
\ref{fig:mmpi0_3pi0},  together  with  the  phase  space  Monte  Carlo
simulations  of $pd  \rightarrow ^3$He$\pi^0\pi^0\pi^0$.  There  is a small
enhancement  at   high  $MM(\pi^0)$  around  the   $2  m_p  +
M_{N^*(1440)}$ sum, which may  indicate the involvement of a Roper $N^*(1440)$ excitation
in the production mechanism.

\subsection{$pd \rightarrow ^3\mathrm{He}\,\pi^+\pi^-\pi^0$}
\label{sec:pipimpi0}
All    selection   criteria    optimised    for   $\eta    \rightarrow
\pi^+\pi^-\pi^0$ selection, given  in table \ref{tab:eta_pipimpi0t} in
section \ref{sec:eta_pipimpi0},  are applied. In  addition, we require
that both charged  pions are emitted in directions  covered by the CD,
\textit{i.e.} that no  other charged tracks than the  $^3$He are found
in the FD.  Finally, two charged tracks in the  MDC are required, with
overlapping hits  in the PSB. The total  acceptance, calculated partly
using Monte Carlo simulations  (all constraints not involving the MDC,
see  section  \ref{sec:eta_pipimpi0})  and  partly  analysing  $\omega
\rightarrow \pi^+\pi^-\pi^0$ data  (all constraints involving the MDC,
see\\ Ref.\cite{karinthesis})  is 7.2\%  at 1360 MeV  and 6.7\%  at 1450
MeV.

\begin{table}
\begin{center}
\caption{The constraints applied for selection of \mbox{$pd \rightarrow ^3\mathrm{He}\,\pi^+\pi^-\pi^0$}}
\begin{tabular}{l}
\hline
 $^3$He giving signal in the FPC and stopping in the FRH \\
 $\geqslant$ 2 photons in the SEC \\
 one $\gamma\gamma$-combination fulfilling \mbox{$|IM(\gamma\gamma)-m_{\pi^0}|< 45~\rm{MeV}/c^2$} \\
 $MM(^3$He$\pi^0)$ $>$ 250 $\rm{MeV}/c^2$ \\
 $\geqslant$ 2 hits in the PSB \\
 $E_{\rm{tot}}(\rm{SEC})$ $<$ 900 MeV \\
 no $\pi^{\pm}$ in FD \\
 2 MDC tracks with matching hits in the PSB \\
 \hline
\end{tabular}
\label{tab:pipimpi0t}
\end{center}
\end{table}

\begin{figure}
\begin{center}
\resizebox{0.5\textwidth}{!}{%
\includegraphics{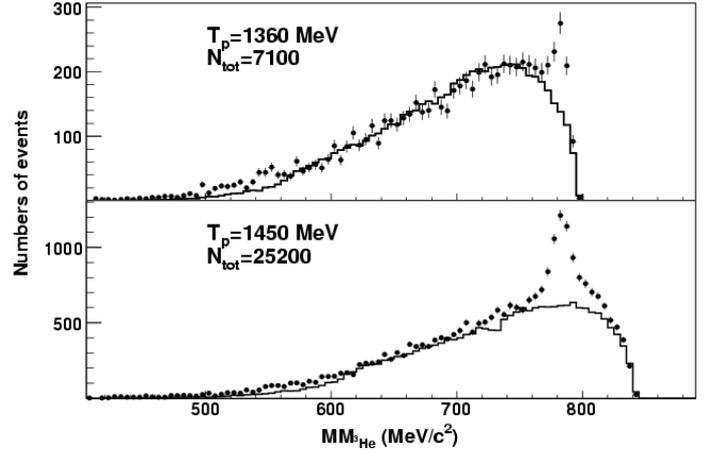}
}
\caption{The  upper panel shows  the WASA  data sample  fulfilling the
  constraints   optimised    for   selection   of    $pd   \rightarrow
  ^3\mathrm{He}\,\pi^+\pi^-\pi^0$  at  1360 MeV  and  the lower  panel
  shows the 1450 MeV case.  The solid line histograms show Monte Carlo
  simulated  $pd  \rightarrow  \mbox{$^3$He$\,\pi^+\pi^-\pi^0$}$  data
  fulfilling the given constraints.  The peak in the experimental data
  at     high      missing     masses     are      $pd     \rightarrow
  ^3\mathrm{He}\,\omega,\,\omega      \rightarrow     \pi^+\pi^-\pi^0$
  events.  The  spectra  are  not  corrected for  acceptance  and  the
  background simulations are scaled to fit the data.}
\label{fig:pipimpi0f}       
\end{center}
\end{figure}

In figure \ref{fig:pipimpi0f}, the missing mass of the $^3$He is shown
for   all   events   satisfying    the   criteria   given   in   table
\ref{tab:pipimpi0t}. There  is a  small enhancement around  the $\eta$
mass and a clear peak at  the $\omega$ mass, but except from that, the
experimental  data seem  to follow  the phase  space $\pi^+\pi^-\pi^0$
distribution  well.    There  is  no  sign  of   any  $pd  \rightarrow
^3$He$\,\eta\pi^0,  \eta \rightarrow  \pi^+\pi^-\pi^0$ events  in this
sample. The  acceptance for  this reaction when  applying the  cuts in
table \ref{tab:pipimpi0t} is 9\% at 1360  MeV and 12\% at 1450 MeV. At
the higher energy,  where the $\eta\pi^0$ cross section  is known (see
\cite{karinetapi}),   the   expected   number   of   $pd   \rightarrow
^3$He$\,\eta\pi^0,   \eta  \rightarrow   \pi^+\pi^-\pi^0$   events  is
$\approx$250, which constitute only $\approx$1\% of the continuum data
in  the lower panel  of figure  \ref{fig:pipimpi0f}.  In  the previous
section we  found that the direct  $4\pi^0$ cross section  at 1450 MeV
must  be  very  small,  and  even  if  the  cross  section  of  direct
$\pi^+\pi^-\pi^0\pi^0$ is likely higher than the direct $4\pi^0$ cross
section due to  more possible isospin amplitudes, it  is reasonable to
assume  that  direct  $\pi^+\pi^-\pi^0\pi^0$  production will  give  a
negligible  contribution to  our $\pi^+\pi^-\pi^0$  data  sample. This
assumption is also  very well in line with  the good agreement between
data  and simulations  in figure  \ref{fig:pipimpi0f}.  The  number of
$\pi^+\pi^-\pi^0$ events  at 1360 MeV, $N_{3\pi}$  = 6700, corresponds
to  a total  cross  section  of 1400  nb. This is obtained using an acceptance
which is calculated assuming phase space production, but since the WASA detector covers 
the major part of the $^3$He phase space for this reaction, the model dependence of the acceptance is small.

The largest contribution to  the systematic uncertainty comes from the
efficiency of the MDC. A  robust method of estimating this uncertainty
is  to  calculate  the  cross  section  with  and  without  using  the
information  from the  MDC,  treat the  difference  as a  systematical
uncertainty and assume that it is symmetric.

At 1360 MeV, the  number  of  $\pi^+\pi^-\pi^0$  events  obtained  using  selection
criteria  without involving  the MDC  is  $N_{3\pi}$ =  33000 and  the
acceptance is  28\%. This  corresponds to a  cross section of  1770 nb
which gives a systematical uncertainty of 370 nb.

There  may   be  a  systematical  uncertainty   arising  from  falsely
identified   $pd  \rightarrow   ^3$He$\,\eta\pi^0,   \eta  \rightarrow
\pi^+\pi^-\pi^0$.  According to  the  rough upper limit  of the  $\eta\pi^0$
cross  section at 1360 MeV that  was given  in the  previous section,  the maximum
number  of $\eta\pi^0$  events  in this sample  is 56.  This gives  a
systematic uncertainty of 12 nb and is thus very small compared to the
uncertainty from the MDC efficiency.

The  statistical uncertainty  is obtained  by the  square root  of the
total number  of events  and is  determined to 17  nb. Finally  we get
$\sigma_{\pi^+\pi^-\pi^0}  = 1400  \pm 17  \pm 370$  nb $\pm  29\%$ at
$T_p$ = 1360 MeV.

Following the same reasoning at  1450 MeV, except that the $\eta\pi^0$
cross section is known with  good precision, the cross section becomes
$\sigma_{\pi^+\pi^-\pi^0} = 910 \pm 7 \pm 80$ nb $\pm 12\%$.

\begin{figure}
\begin{center}
\includegraphics[width=\columnwidth,clip]{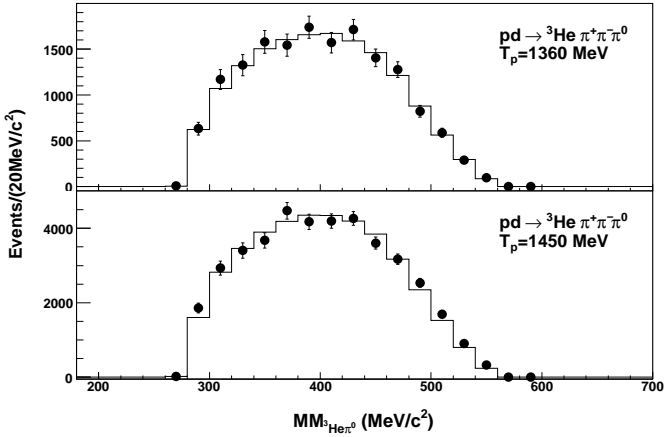}
\caption{The  missing  mass of  the  $^3$He$\pi^0$-system. The  points
  represent acceptance corrected data satisfying the criteria given in
  the text and $600$ MeV$/c^2 < MM(^3$He$) < 700$MeV $/c^2$. The solid
  histogram   shows   phase   space   Monte   Carlo   simulations   of
  $\pi^+\pi^-\pi^0$ production. The upper panel shows the $T_p$ = 1360
  MeV case and the lower the $T_p$ = 1450 MeV case.}
\label{fig:im2p_pipimpi0}       
\end{center}
\end{figure}

As  in  the  $3\pi^0$  case,  the distributions  of  the  final  state
particles have been  studied. Since the $\pi^0$ is  the only pion that
is fully reconstructed, the  invariant mass of the $\pi^+\pi^-$-system
is   studied    by   reconstructing   the   missing    mass   of   the
$^3$He$\pi^0$-system. To have a sample as similar to the $3\pi^0$ case
as possible, events  which satisfy $600$ MeV$/c^2 <  MM(^3$He$) < 700$
MeV$/c^2$ are selected. The $MM(^3$He$\pi^0)$ is reconstructed and the
data are corrected for acceptance. Note that no background subtraction
had to be made in this case, since the contribution from $pd  \rightarrow   ^3$He$\,\eta\pi^0$
 is proven to be small at both energies. The results at both energies are shown in
figure  \ref{fig:im2p_pipimpi0}.  There   is  good  agreement  between
experiment and  simulated $\pi^+\pi^-\pi^0$ data and there  is no sign
of   any   intermediate   $\rho$   meson,   which   would   push   the
$MM(^3$He$\pi^0)$ towards higher masses.  This is not surprising since
despite  the large  width of  the  $\rho$ meson  ($\Gamma \approx$  150
MeV), we are far below the nominal $pd \rightarrow ^3$He$\rho^0\pi^0$
threshold at both beam energies considered in this work.

\begin{figure}
\begin{center}
\includegraphics[width=\columnwidth,clip]{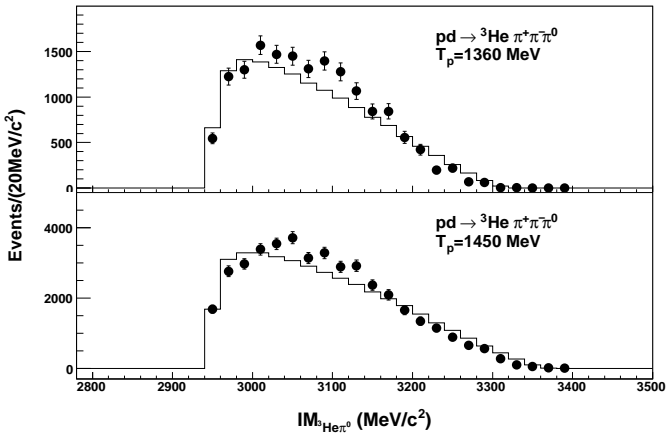}
\caption{The  invariant mass of  the $^3$He$\pi^0$-system.  The points
  represent acceptance corrected data fulfilling the constraints given
  in the  text and $600$ MeV$/c^2  < MM(^3$He$) <  700$ MeV$/c^2$. The
  solid  histogram  shows  phase  space  Monte  Carlo  simulations  of
  $\pi^+\pi^-\pi^0$ production. The upper panel shows the $T_p$ = 1360
  MeV case and the lower the $T_p$ = 1450 MeV case.}
\label{fig:imhp_pipimpi0}       
\end{center}
\end{figure}

The invariant mass  of the $^3$He$\pi^0$ is reconstructed  in the same
way as in the $3\pi^0$ case, except that no background subtraction was
necessary.      The     results      are      shown     in      figure
\ref{fig:imhp_pipimpi0}. Also in  the $\pi^+\pi^-\pi^0$ case, there is
a small enhancement around the $2 m_p + M_{\Delta(1232)}$ sum.

\begin{figure}
\begin{center}
\includegraphics[width=\columnwidth,clip]{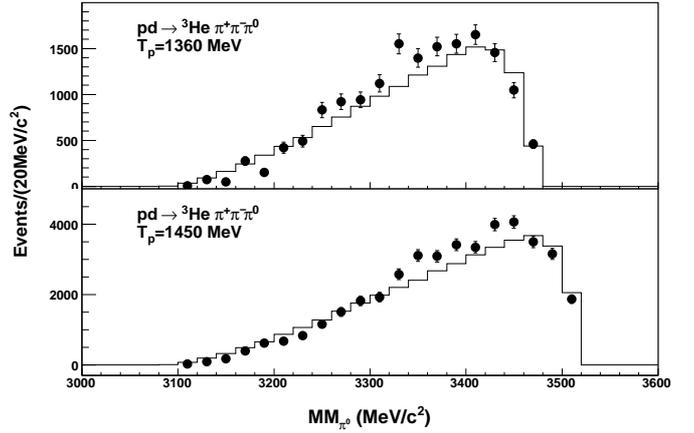}
\caption{The  missing  mass  of  the  $\pi^0$.  The  points  represent
  acceptance corrected  data fulfilling  the constraints given  in the
  text and  $600$ MeV$/c^2  < MM(^3$He$) <  700$ MeV$/c^2$.  The solid
  histogram   shows   phase   space   Monte   Carlo   simulations   of
  $\pi^+\pi^-\pi^0$ production. The upper panel shows the $T_p$ = 1360
  MeV case and the lower the $T_p$ = 1450 MeV case.}
\label{fig:mmpi0_pipimpi0}       
\end{center}
\end{figure}

The  invariant mass  of the  $^3$He$\pi^+\pi^-$ system  is  studied by
reconstructing  the  missing  mass  of the  $\pi^0$.   The  acceptance
corrected data  are shown in figure  \ref{fig:mmpi0_pipimpi0}. Like in
the $pd  \rightarrow ^3$He$\,\pi^0\pi^0\pi^0$  case, there is  a small
enhancement   with  respect   to  phase   space  near   the   $2m_p  +
M_{N^*(1440)}$ sum, suggesting that the Roper resonance $N^*(1440)$ may be
involved in the production process.

\subsection{Comparison between the $pd \rightarrow ^3\mathrm{He}\,\pi^+\pi^-\pi^0$ and the $pd \rightarrow ^3\mathrm{He}\,\pi^0\pi^0\pi^0$ reactions}
\label{sec:ratio}

In the introduction,  the ratio between the cross  sections of the $pd
\rightarrow  ^3\mathrm{He}\,\pi^+\pi^-\pi^0$ and  the  $pd \rightarrow
^3\mathrm{He}\,\pi^0\pi^0\pi^0$                              reactions,
\textit{i.e.}\\$\frac{\sigma(pd                             \rightarrow
  ^3\mathrm{He}\,\pi^+\pi^-\pi^0)}{\sigma(pd                \rightarrow
  ^3\mathrm{He}\,\pi^0\pi^0\pi^0)}$,  was  calculated  to 9,  using  a
statistical model where all  isospin amplitudes $M_{T_{3\pi}}$ are put
equal and all cross terms are set to zero. If instead $M_{0}$ = 0, the
ratio  becomes  4.   In  this   work,  the  ratio  has  been  measured
experimentally  at  both  energies.    By  using  the  cross  sections
determined in section  \ref{sec:3pi0} and \ref{sec:pipimpi0}, one then
obtains 7.8 at  1360 MeV and 7.9 at 1450 MeV  for this ratio. However,
to give a  comparison at the same excess energy  $Q$, the results have
to  be  corrected  for  the  difference  between  the  masses  of  the
$\pi^{\pm}$ and $\pi^0$. The lower mass of the $\pi^0$ makes the phase
space volume of  the $pd \rightarrow ^3$He$\,\pi^0\pi^0\pi^0$ reaction
larger      than      that       of      the      $pd      \rightarrow
^3\mathrm{He}\,\pi^+\pi^-\pi^0$  at  the   same  beam  energy.   After
correcting for the difference in phase space volume, the ratio becomes
$8.3 \pm 0.3  \pm 3.1$ at 1360 MeV  and $8.4 \pm 0.2 \pm  1.8$ at 1450
MeV,  where  the  first   uncertainty  is  statistic  and  the  second
systematic. Note that the  uncertainty in the normalisation cancels in
the  ratio. The values  obtained are  consistent with  the value  of 9
predicted using  the statistical approach. The  interpretation of this
result is then that $M_{0}$ should be of similar size as $M_{1}$.

\section{Summary and Conclusions}
\label{sec:sumall}
The production  of light mesons, \textit{i.e.} $\pi$  and $\eta$, have
been  studied at  $T_p$  = 1360  MeV and  $T_p$  = 1450  MeV. The  $pd
\rightarrow ^3$He$\,\eta$ reaction was  studied by using data from the
three  most common  decay channels;  $\eta  \rightarrow \gamma\gamma$,
$\eta    \rightarrow    \pi^0\pi^0\pi^0$    and   $\eta    \rightarrow
\pi^+\pi^-\pi^0$.  The   result  from  the   different  channels  gave
consistent results. At both energies, the angular distributions of the
$\eta$   meson  were   reconstructed  and   they  show   a  pronounced
forward-backward asymmetry. The WASA  detector does not cover the full
angular range at  these high energies and one  could therefore not say
whether the forward plateau  or dip observed in \cite{bilger,rausmann}
persists at high energies. The data from this work in combination with
data  taken  at  the  same  energy  with  the  SPES  III  spectrometer
\cite{Kirchner} do, however, suggest  that the forward dip persists at
this high energy and is stronger  compared to the lower energy case. If not, the
1450 MeV  data from this work disagree  with the SPES III  data in the
forward hemisphere.

The total cross  sections of three pion production  in\\ $pd \rightarrow
^3$He$\,\pi^0\pi^0\pi^0$ and  $pd \rightarrow ^3$He$\,\pi^+\pi^-\pi^0$
were   measured   at   both    energies.   The   ratio   between   the
$\pi^+\pi^-\pi^0$ and  $3\pi^0$ cross sections was  calculated at both
energies and  the results are  consistent with the  statistical model,
where $M_{0} = M_{1}$ and all cross terms are neglected. The invariant
mass distributions of the two-pion system and the $^3$He$\pi^0$ system
were reconstructed.  The invariant mass distributions  of the two-pion
system follow  phase space but  the corresponding distribution  of the
$^3$He$\pi^0$  system, show  a small  enhancement around  the  $2m_p +
M_{\Delta(1232)}$ mass. This enhancement  was observed in $3\pi^0$ and
$\pi^+\pi^-\pi^0$  production  at both  energies  and  may indicate  a
single $\Delta$ excitation in the production mechanism.

The invariant mass distributions  of the $^3$He$2\pi$ system, for both
$\pi^+\pi^-\pi^0$ and  $3\pi^0$ production at both  eneregies, show an
enhancement  near  the $2m_p  +  M_{N^*(1440)}$  mass, suggesting  that the  Roper $N^*(1440)$  resonance may be involved in  the production
mechanism.

The  cross sections  measured in  this  work are  summarised in  table
\ref{tab:all_crossec}.

\begin{table}
\begin{center}
\caption{The  total cross sections  of the  reactions studied  in this
  work. In addition to the  systematic uncertainty given in the table,
  there is a normalisation uncertainty of 29\% at 1360 MeV and 12\% at
  1450  MeV.   The  normalisation   is  made  using   $pd  \rightarrow
  ^3$He$\,\eta$  data   and  differential  cross   sections  given  in
  \cite{Berthet,Kirchner}.}
\begin{tabular}{lllll}
\hline
Reaction & $T_p$ (MeV) & $\sigma$ & stat. & syst.\\
  & & ($nb$) & ($nb$) & ($nb$) \\
\hline
$pd \rightarrow ^3$He$\,\eta$ & 1360 & 151.6  & $\pm$ 9.3  & $\pm$ 35.2$$ \\
$pd \rightarrow ^3$He$\,\pi^0\pi^0\pi^0$ & 1360 & 180 & $\pm$ 6 & $\pm 49$ \\
$pd \rightarrow ^3$He$\,\pi^+\pi^-\pi^0$ & 1360 & 1400 & $\pm$ 17 & $\pm 370$ \\
\hline
$pd \rightarrow ^3$He$\,\eta$ & 1450 & 80.8  & $\pm$ 3.6  & $\pm$ 43.1$$ \\
$pd \rightarrow ^3$He$\,\pi^0\pi^0\pi^0$ & 1450 & 115 & $\pm$ 3 & $\pm 23$ \\
$pd \rightarrow ^3$He$\,\pi^+\pi^-\pi^0$ & 1450 & 910 & $\pm$ 7 & $\pm 80$ \\
\hline
\end{tabular}
\label{tab:all_crossec}
\end{center}
\end{table}

\section{Acknowledgements}
We are  grateful to the personnel  at the The  Svedberg Laboratory for
their  support during  the course  of  the experiment.  This work  was
supported  by  the  European  Community under  the  ``Structuring  the
European Research  Area'' Specific Programme  research Infrastructures
Action (Hadron Physics, contract number RII3-cT-204-506078) and by the
Swedish Research Council.

%

%
%

\end{document}